**Neural correlates of self-generated imagery and cognition throughout the sleep cycle**

Kieran C. R. Fox[a,*] and Manesh Girn[a]

[a] Department of Psychology, University of British Columbia, 2136 West Mall, Vancouver, B.C., V6T 1Z4 Canada

[*] To whom correspondence may be addressed (at address [a] above). Telephone: 1-778-968-3334; Fax: 1-604-822-6923; E-mail: kfox@psych.ubc.ca

**Abstract**

Humans have been aware for thousands of years that sleep comes in many forms, accompanied by different kinds of mental content. We review the first-person report literature on the frequency and type of content experienced in various stages of sleep, showing that different sleep stages are dissociable at the subjective level. We then relate these subjective differences to the growing literature differentiating the various sleep stages at the neurophysiological level, including evidence from electrophysiology, neurochemistry, and functional neuroimaging. We suggest that there is emerging evidence for relationships between sleep stage, neurophysiological activity, and subjective experiences. Specifically, we emphasize that functional neuroimaging work suggests a parallel between activation and deactivation of default network and visual network brain areas and the varying frequency and intensity of imagery and dream mentation across sleep stages; additionally, frontoparietal control network activity across sleep stages may parallel levels of cognitive control and meta-awareness. Together these findings suggest intriguing brain-mind isomorphisms and may serve as a first step toward a comprehensive understanding of the relationship between neurophysiology and psychology in sleep and dreaming.

**Introduction: The multiplicity of sleep**

Awareness of the subjective multiplicity of sleep goes back thousands of years – at least as far as the ancient Indian philosophical texts known as the *Upanishads* composed around the 6[th] century B.C.E. (Deutsch & Dalvi, 2004; Hume, 1921; Prabhavananda, Manchester, & Isherwood, 1984; Sharma, 2012). Ancient Indian philosophers clearly recognized a distinction between dreamless sleep, dreaming, and even 'lucid' dreaming – being aware that one is dreaming while dreaming (Prabhavananda et al., 1984; Sharma, 2012). In the West, Aristotle made strikingly prescient observations for his time: he recognized both dreamless and dreaming sleep; described what we today call sleep-onset hypnagogic imagery; correctly hypothesized that dreaming represents the activity of our perceptual faculties in the absence of external inputs; and even recognized the possibility of lucid dreaming (Aristoteles & Gallop, 1996; Barbera, 2008). And at least one thousand years ago, Tibetan Buddhist practitioners had developed sophisticated cognitive practices geared toward increasing metacognitive awareness during dreamless sleep and dreaming (Gillespie, 1988; Wangyal, 1998). These traditions began a fruitful mapping of quantitative and qualitative psychological ($\psi$) differences throughout the sleep cycle.

Western science began to finally put these observations on a firmer footing in the mid-20[th] century, with the discovery that surface-recorded brain electrical potentials could dissociate between several sleep stages (Aserinsky & Kleitman, 1953, 1955; Dement & Kleitman, 1957a, 1957b; Monroe, Rechtschaffen, Foulkes, & Jensen, 1965). This research, the first to definitively identify neurobiological ($\Phi$) markers related to particular cognitive states ($\psi$) during sleep, led to the well-known classification of sleep into rapid-eye-movement (REM) and non-rapid-eye-movement (NREM) stages (Rechtschaffen & Kales, 1968), with four major stages generally recognized by contemporary researchers (NREM1, 2, 3/4, and REM).

These stages have not been equally recognized or researched over the past few decades. REM and NREM (the latter of which had not yet been differentiated into substages, and was generally known as 'slow wave sleep' – SWS) were intensively investigated from the beginning of modern sleep and dream science in the 1950s. It was rapidly recognized that these stages were characterized by differences in subjective experience – most notably, by the high frequency of dream reports following awakening from REM, but the relative paucity of such reports following awakenings from NREM (reviewed by Nielsen, 1999, 2000).

A pair of more marginal and difficult to investigate stages were largely ignored until relatively recently: NREM1 (sleep onset) and so-called 'lucid dreaming,' in which one is aware of the fact that one is dreaming *while* dreaming (LaBerge, Nagel, Dement, & Zarcone Jr, 1981). Detailed investigation of the electrophysiological substages and phenomenological content of sleep onset (NREM1), although inaugurated in the 1960s (Foulkes, Spear, & Symonds, 1966; Foulkes & Vogel, 1965; Vogel, Foulkes, & Trosman, 1966), was not conducted in earnest until the 1990s (Hayashi, Katoh, & Hori, 1999; Hori, Hayashi, & Morikawa, 1994; Tanaka, Hayashi, & Hori, 1996, 1997). Lucid REM sleep dreaming (LREM) still remains controversial to many researchers; pioneering but tenuous polysomnographic research from the 1980s (Fenwick et al., 1984; LaBerge, 1980; LaBerge, Levitan, & Dement, 1986; LaBerge et al., 1981) has continued to be replicated and extended, however (Holzinger, LaBerge, & Levitan, 2006; LaBerge & Levitan, 1995; Voss, Holzmann, Tuin, & Hobson, 2009), as well as investigated with more sophisticated methods, such as combined EEG-fMRI (Dresler et al., 2011; Dresler et al., 2012) and transcranial direct current stimulation (Stumbrys, Erlacher, & Schredl, 2013).

In this chapter, we aim to briefly review what is known about the differentiability of sleep stages according to various neurophysiological methods and markers, and to relate these neurophysiological differences to variations in subjective experience across the sleep cycle as indicated by first-person reports.

*Sleep mentation as self-generated thought*

We and others have argued at length elsewhere that mentation during sleep, particularly dreaming *per se*, can be viewed as an intensified form of waking self-generated thought or mind-wandering (Christoff, Irving, Fox, Spreng, & Andrews-Hanna, 2016; Domhoff, 2011; Domhoff & Fox, 2015; Fox, Nijeboer, Solomonova, Domhoff, & Christoff, 2013); see also Domhoff, this volume). The basis for this claim is twofold: both the subjective experience of dreaming, and its neurophysiological correlates (as indexed by REM sleep), parallel those of waking mind-wandering and related forms of self-generated thought.

Waking self-generated thought is typically characterized by auditory and visual imagery, ubiquitous affect, a strong focus on current concerns and social interactions, and varying degrees of narrative structure (Andrews-Hanna, Smallwood, & Spreng, 2014; Fox et al., 2013; Fox, Spreng, Ellamil, Andrews-Hanna, & Christoff, 2015); see also Stawarczyk, this volume). The same statements can be made about REM sleep mentation, with the qualification that these characteristics in fact tend to be heightened or exaggerated in dreaming: the audiovisual world is fully present and immersive, emotions are more intense and perhaps more ubiquitous, social characters are more numerous and interactions with them more elaborate, and narrative structure is extended over time and in more complex ways (Domhoff & Fox, 2015; Fox et al., 2013; Windt, 2010).

A similar parallel is observed at the neurophysiological level. Waking self-generated thought, as compared to active focus on a task or external stimulus, is associated with a relatively consistent pattern of brain activations centered on the default network and extending into medial occipital areas involved in visual imagery, as well as some executive brain regions tied to the frontoparietal control network (Fox et al., 2015). When our group meta-analyzed functional neuroimaging studies of REM sleep, during which dreaming occurs approximately 80% of the time (Hobson, Pace-Schott, & Stickgold, 2000), we found that many of the same brain areas implicated in waking self-generated thought were even more strongly recruited during REM sleep, including medial prefrontal cortex, numerous medial temporal lobe structures, and medial occipital areas (Fox et al., 2013). Additionally, by slightly relaxing statistical thresholds, further overlap in the inferior parietal lobule, another key default network region, was revealed (Domhoff & Fox, 2015).

Overall, these results suggested to us that dreaming, and its most common neurophysiological correlate, REM sleep, show an overall intensification or amplification of both the subjective qualities and neural recruitment associated with waking self-generated cognition (cf. Fig. 3 in Fox et al., 2013). Due to the fact that the NREM sleep stages are also characterized by variable levels of cognitive activity and dream experience, determining their general neural correlates presents an attractive target for research: a general understanding of these neural substrates would allow further examination of the hypothesis that self-generated thought has a common brain basis independent of the particular conscious state in which it takes place. Reviewing the general neural correlates of the NREM sleep stages, and how they might fit into the spectrum of self-generated cognition across wake and dreaming, is therefore the main aim of this chapter.

*Sleep can be meaningfully dissociated into stages*

The preceding overview only hints at the enormous body of work that has been conducted over the past 60 years within a paradigm whose core assumption is that sleep stages can be meaningfully dissociated and more-or-less independently investigated. Loomis and colleagues (Loomis, Harvey, & Hobart, 1935; Loomis, Harvey, & Hobart, 1937) were the first to provide detailed descriptions of distinct neurophysiological stages in normal human sleep, and much subsequent work has followed, corroborated, and expanded on these efforts. What are the criteria upon which these sleep stages are distinguished – and are they valid?

Nielsen (2014) rightly points out that the widespread use of standard sleep stage scoring criteria (Rechtschaffen & Kales, 1968) has led to an artificially categorical view of sleep stages, accompanied by tacit assumptions of both mutual exclusivity and abrupt transitions. Even the narrow use of just a few electrophysiological markers cannot support such a view; and of the hundreds of potential physiological and neural markers that fluctuate throughout the sleep-wake cycle, only a select few are routinely employed to differentiate sleep stages (Nielsen, 2014). These facts should put us on guard against any facile reification of distinct sleep stages.

A key question therefore needs to be asked: Are sleep stages a fact of neurophysiology or an investigative convenience? The answer, we believe, is that they are somewhere in between. While keeping the above caveats firmly in mind and refraining from reifying classification schemes as actual entities, we agree with the conclusions of various comprehensive reviews of this issue (e.g., (Hobson et al., 2000; Nielsen, 2000): persuasive evidence argues for the distinctiveness of sleep stages *in general*. Although the various major sleep stages share features in common, can oscillate back and forth unpredictably, and may hybridize and give rise to not-easily-classified transitional stages (Nielsen, 2014), meaningful (if tentative) statements can nonetheless be made about their characteristic patterns of phenomenology, electrophysiology, and neurochemistry (Hobson et al., 2000; Nielsen, 2000).

In the following sections we present these multiple lines of evidence in support of the utility and plausibility of distinctive (if not entirely mutually exclusive) sleep stages. We argue that these neurophysiological and phenomenological idiosyncrasies lead to the strong and testable hypothesis that patterns of brain activation, as measured by relatively non-invasive functional neuroimaging methods like fMRI and positron emission tomography (PET), should vary accordingly across the NREM and REM sleep stages. Moreover, the finding that REM sleep (with high chances of dreaming) shares many neural correlates with waking self-generated thought (Domhoff, 2011; Domhoff & Fox, 2015; Fox et al., 2013), coupled with the knowledge that self-generated thought frequency and vividness differ markedly across sleep stages (Hobson et al., 2000; Nielsen, 2000), leads to the even stronger hypothesis that specific brain areas involved in self-generated thought should show corresponding activation and deactivation throughout the sleep cycle in concert with subjectively-experienced differences in content.

The body of this chapter provides an overview of all neuroimaging studies of sleep and dreaming in humans, in an effort to synthesize what has been learned from two decades of investigations of brain activations and deactivations throughout the sleep cycle. The aim is not to argue for strict one-to-one isomorphisms between mental states ($\psi$) and neuromarkers ($\Phi$), but rather to summarize the current evidence for broad and intriguing correspondences between first- and third-person levels of description.

## General evidence for psychological and neurophysiological differences across sleep stages

*Phenomenological dissociation of sleep stages*

The subjectively-experienced ('phenomenological' for our purposes) differences in the experience of sleep stages have been noted for millennia (Aristoteles & Gallop, 1996) (Deutsch & Dalvi, 2004; Sharma, 2012; Thompson, 2014), but systematic research using large, representative samples has taken place mostly in the past few decades (Domhoff, 2003; Nielsen, 2000). Although many methodological difficulties (and almost as many theoretical deadlocks) have burdened this otherwise burgeoning field, some general conclusions can be cautiously drawn regarding differences in the frequency, quality, and content of mentation across sleep stages.

Nielsen (1999; 2000) has thoroughly summarized this literature, highlighting the critical distinction between *cognitive activity* in general (which can include thought-like mentation, isolated flashes of imagery, and so on) and truly immersive and hallucinatory *dreaming* proper (a particular subset of cognitive activity). Whereas cognitive activity is fairly prevalent throughout all sleep stages (at least 40% of awakenings from any given sleep stage will lead to a report of some kind of cognitive activity; Nielsen, 1999), dreaming proper is largely restricted to REM sleep and certain sub-stages of NREM1 sleep onset, and is comparatively rare during other NREM sleep stages (Table 1).

**Table 1.** Approximate frequency of subjective reports of cognitive activity and dreaming across sleep stages.

| Sleep Stage | Cognitive Activity | Dreaming | Predominant EEG rhythm |
|---|---|---|---|
| **NREM1 (sleep onset)** | ~40% | ~35% | Alpha, Theta |
| **NREM2** | 50% | ~15-25% | Theta, Spindles (Beta) |
| **NREM3/4 (SWS)** | 40-50% | ~10% | Delta |
| **REM** | 80% | 80% | Theta, Beta |
| **LREM** | 100% | 100% | Theta, Beta, Gamma |

EEG: electroencephalography; LREM: lucid REM; NREM: non-rapid eye movement; REM: rapid eye movement; SO: sleep onset; SWS: slow-wave sleep. Table based upon comprehensive reviews by Nielsen (1999; 2000) and Hobson et al. (2000).

*Electrophysiological dissociation of sleep stages*

The first and best-known neurophysiological (Φ) division of sleep is based on scalp electrode recordings of pooled neuronal electrical potentials, presumed to represent primarily the summation of post-synaptic potentials throughout dendritic arbors and cell somata, and to a lesser extent, synchronous discharge (action potential firing) of populations of neurons (Buzsáki, Anastassiou, & Koch, 2012; Olejniczak, 2006).

The central findings regarding EEG correlates of sleep are summarized in Table 1 and Figure 1. Briefly, resting (eyes-closed) wakefulness is characterized by alpha rhythms (8-12 Hz); the transition to NREM1 (sleep onset) is defined by the gradual disappearance of alpha and the appearance of theta (4-7 Hz) ripples and rolling eye movements. NREM2 begins when high-frequency spindles (in the beta frequency; 12.5-30 Hz) and large-amplitude K-complexes appear frequently in the EEG. NREM3/4 or SWS is characterized instead by very slow delta band (0.5-4 Hz) activity, synchronized in large-amplitude waves. Finally, REM sleep involves a return to highly desynchronized and low amplitude activity predominantly in the theta and beta bands, similar to active wakefulness. Lucid REM sleep, in the few investigations so far conducted, involves an EEG pattern similar to REM sleep but with increased power in the gamma (>30 Hz) band (Voss et al., 2014; Voss et al., 2009). As these electrophysiological differences are well-validated and expertly reviewed elsewhere (Antrobus, 1991; Silber et al., 2007; Williams, Karacan, & Hursch, 1974), they are not discussed further here.

| Stage | sub stages[1] | EEG Signature | EEG Signature Waveform |
|---|---|---|---|
| Wake | 1 | Alpha wave train | 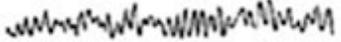 |
| | 2 | Alpha wave intermittent (>50%) | 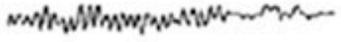 |
| NREM 1 | 3 | Alpha wave intermittent (<50%) | 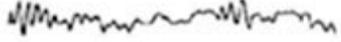 |
| | 4 | EEG flattening (<20 µV) | 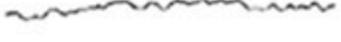 |
| | 5 | Theta ripples | 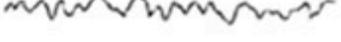 |
| | 6 | Vertex sharp wave (<200 µV) | 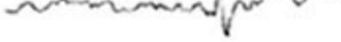 |
| | 7 | > 1 Vertex sharp wave (<200 µV) | 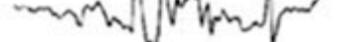 |
| | 8 | Incomplete spindle | 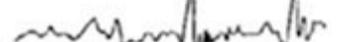 |
| NREM 2 | 9 | Complete Spindle | 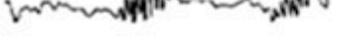 |
| NREM 3-4 | | Delta | 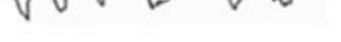 |
| REM | | Theta | 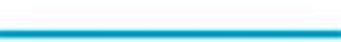 |

**Figure 1.** Main electrophysiological correlates of each sleep stage. NREM1 sub-stages are included, as is eyes-closed waking rest, for comparison. Note the easily-differentiable EEG signature accompanying each sleep stage. Reproduced from Stenstrom, Fox, et al. (2012).

*Neurochemical dissociation of sleep stages*

The patterns of neurochemical activity throughout the sleep-wake cycle are exceedingly complex and consequently poorly understood. At least a dozen major neurotransmitters are involved in regulating NREM and REM, by virtue of either increased or decreased activity (compared to waking) during various sleep stages, but many secondary players with less clear roles are also involved (Gottesmann, 1999; Hobson et al., 2000; Kahn, Pace-Schott, & Hobson, 1997; Lena et al., 2005; E. F. Pace-Schott & Hobson, 2002; Stenberg, 2007). Moreover, changes in neurotransmitter levels are far from a uniform phenomenon throughout the brain: region-by-transmitter interactions have in some cases been experimentally demonstrated (e.g., for dopamine: (Lena et al., 2005)), and it seems likely that activity levels for other neurotransmitters will also vary based not just on sleep stage, but also which region of the brain is being investigated. These complexities are further exacerbated by the existence of many receptor subtypes for each neurotransmitter, and the concomitant (and often unknown – at least for sleep) differences in downstream effects caused by the actions of a single neurotransmitter (Monti & Monti, 2007). A final and major difficulty is that implanting recording electrodes directly into the subcortical nuclei responsible for manufacturing and/or disseminating these neurotransmitters is the ideal method of investigation – an approach typically precluded in humans. Almost everything that is known about the neurochemistry of sleep has therefore been drawn from studies of animals with variable phylogenetic proximity to humans, such as rats, cats, rabbits, and monkeys (Gottesmann, 1999; Jones, 1991, 2005; Lena et al., 2005; Stenberg, 2007).

The enormous difficulty of studying the neurochemistry of sleep and dreaming have so far precluded the formulation of a clear model of each neurotransmitter's relative activity throughout NREM and REM sleep stages, much less what the functional implications of such neurochemical heterogeneity might be. Nonetheless, decades of research have yielded some broad trends, which we summarize in Table 2.

Although a stage-by-stage model is premature given the current limits of our knowledge, broad trends can distinguish waking at least from REM and NREM sleep. Generally speaking, all major neurotransmitters show some level of tonic activity during waking; conversely, all of these neurotransmitters show a greater or lesser decrease of activity during various NREM stages of sleep (it should be noted, however, that much of this data is derived only from the later stages of 'slow-wave' NREM sleep). Finally, REM sleep shows an intermediate pattern: most neurotransmitter activity is decreased relative to waking, but notably, acetylcholine and dopamine levels appear to be elevated (Table 2).

**Table 2.** Neurochemical profiles of the various stages of sleep as compared to waking.

| Neurotransmitter | State | |
|---|---|---|
| | REM (↑ Dreaming) | NREM (↓ Dreaming) |
| ACh | ↑↑ | ↓ |
| Asp | ↓ | ↓↓ |
| DA | ↑↑ | ↓ |
| GABA | **?** | **?** |
| Glu | ↓↓ | ↓ |
| HA | ↓↓ | ↓ |
| NE | ↓↓ | ↓ |
| 5-HT | ↓↓ | ↓ |

ACh: acetylcholine; Asp: aspartate; DA: dopamine: GABA: $\gamma$-amino butyric acid; Glu: glutamate; HA: histamine; NE: norepinephrine/noradrenaline; NREM: non-rapid-eye-movement sleep; REM: rapid-eye-movement sleep; 5-HT: serotonin; SWS: slow wave sleep. Based on data from (Gottesmann, 1999; Hobson, 2009; Lena et al., 2005; E. F. Pace-Schott & Hobson, 2002).

**Functional neuroimaging of sleep and dreaming: A fourth line of evidence**

*Overview*

The preceding sections have highlighted some of the key evidence supporting the hypothesis that stages of sleep are dissociable in terms of their (i) phenomenology, (ii) electrophysiology, and more provisionally, (iii) neurochemistry. These many heterogeneities across sleep stages lead to the strong and testable hypothesis that brain activations and deactivations, as detected with non-invasive functional neuroimaging modalities, might also show dissociable patterns throughout the sleep cycle. Functional neuroimaging therefore provides a fourth line of evidence that could either corroborate (or possibly contradict) the ample body of research suggesting phenomenological and neurophysiological dissociability. As noted above, the phenomenological variations in self-generated thought across sleep stages further suggest that differential activation patterns, if they exist, should exhibit some relationship with the numerous regions implicated in self-generated thought in waking (Fox et al., 2015) and REM sleep (Fox et al., 2013).

Despite a profusion of neuroimaging studies elucidating the neural basis of sleep and dreaming (Desseilles, Dang-Vu, Sterpenich, & Schwartz, 2011; Maquet, 2010; Edward F Pace-Schott, 2007), no comprehensive overview has been conducted in recent years. While distinctive patterns of observed brain activity appear to largely parallel the differential subjective content reported from laboratory awakenings across the various sleep stages, there has yet to be any systematic attempt to specifically relate these particular patterns of brain activity to differing first-person experiences. Recently, we conducted such a meta-analysis and review of subjective content for REM sleep and dreaming (Fox et al., 2013); here, we expand upon those results to include functional neuroimaging and subjective report data from all sleep stages, in a preliminary attempt to relate subjective experience, electrophysiology, neurochemistry, and brain blood-flow-related activity across each sleep stage.

*Literature review*

In order to ensure that our review was comprehensive, we scoured Google Scholar, PubMed, and PsycInfo online databases for any study whose title or abstract included keywords such as 'PET,' 'fMRI,' and 'neuroimaging,' in combination with keywords such as 'sleep,' 'dreaming,' 'NREM,' or 'REM.' The reference list of each study found was also consulted, as was the bibliography of a major review of functional neuroimaging of sleep (Hobson et al., 2000). Our search yielded 58 functional neuroimaging (PET or fMRI) studies of REM and NREM sleep. We limit our discussion here to studies that employed a baseline of resting wakefulness (either pre- or post-sleep), in order to minimize the confounding effects of various tasks and baseline conditions; and to studies that involved neurologically and psychiatrically healthy participants. We also avoid discussing studies that introduced extraneous factors (e.g., auditory stimulation during sleep) or pharmacological agents. A total of 16 studies were ultimately consulted (Table 3), many of which examined more than a single sleep stage.

**Table 3.** Functional neuroimaging studies of sleep reviewed.

| Study | Modality | *N* | Stage(s) investigated | Sleep deprivation? |
|---|---|---|---|---|
| Maquet et al. (1996) | PET | 11 | REM | Y |
| Braun et al. (1997) | PET | 37 | SWS, REM | Y |
| Maquet et al. (1997) | PET | 11 | SWS | Y |
| Nofzinger et al. (1997) | PET | 6 | REM | N |
| Braun et al. (1998) | PET | 10 | REM | Y |
| Kajimura et al. (1999) | PET | 18 | SWS | Y |
| Finelli et al. (2000) | PET | 8 | REM | Y |
| Maquet et al. (2000) | PET | 5 | REM | N |
| Peigneux et al. (2001) | PET | 12 | REM | N |
| Balkin et al. (2002) | PET | 27 | NREM2 | Y |
| Kjaer et al. (2002) | PET | 8 | NREM1 | N |
| Maquet et al. (2005) | PET | 22 | SWS, REM | N |
| Kaufmann et al. (2006) | fMRI | 9 | NREM1, NREM2, SWS | Y |
| Picchioni et al. (2008) | fMRI | 4 | NREM1 | N |
| Andrade et al. (2011) | fMRI | 25 | NREM1, NREM2, SWS | N |
| Koike et al. (2011) | fMRI | 12 | NREM2, SWS, REM | N |

fMRI: functional magnetic resonance imaging; N: no; NREM: non-rapid eye movement sleep; PET: positron emission tomography; REM: rapid eye movement sleep; SWS: slow wave sleep; Y: yes.

*Neural correlates of NREM1 sleep*

As discussed above, NREM1 sleep is a highly heterogeneous sleep stage that can be divided into various sub-stages with varying degrees of visual imagery and cognitive activity. These fine-scale subdivisions on short timescales (or the order of seconds or tens of seconds) mean that is not possible for present functional neuroimaging technologies, with their generally poor temporal resolution, to adequately resolve NREM1 sub-stages.

Nonetheless, a handful of studies (Andrade et al., 2011; Kaufmann et al., 2006; Kjaer, Law, Wiltschiøtz, Paulson, & Madsen, 2002; Picchioni et al., 2008) have investigated the neural correlates of NREM1 (broadly defined) with some intriguing preliminary results. For instance, the first PET study of NREM1 found that, compared to a baseline of resting wakefulness, NREM1 sleep showed greater activation in numerous visual areas, including the fusiform gyrus (BA 19), and the middle occipital gyrus bilaterally (BA 18/19) (Kjaer et al., 2002). Another study also found evidence for medial occipital activation in the cuneus (Kaufmann et al., 2006). Functional connectivity analyses, using the hippocampus as a seed region, have also found increased connectivity with various visual regions, including the fusiform gyrus and the middle and superior occipital gyri (Andrade et al., 2011). Most studies of NREM1 to date therefore show evidence for increased activation of, or coupling with, widespread visual regions (although for an exception to this trend, see the results of Picchioni et al., 2008).

Concurrent with this tentative evidence for visual cortical activation in NREM1, most studies have found evidence for deactivation of prefrontal executive regions, including in superior frontal gyrus (BA 6) (Kjaer et al., 2002; Picchioni et al., 2008) and middle frontal gyrus (BAs 9 and 10) (Kaufmann et al., 2006; Picchioni et al., 2008).

*Neural correlates of NREM2 sleep*

Similar to NREM1, only a small handful of studies have investigated 'pure' NREM2, (Andrade et al., 2011; Balkin et al., 2002; Kaufmann et al., 2006; Koike, Kan, Misaki, & Miyauchi, 2011). Unlike NREM1, however, a general pattern of activations is less easily discernible. For instance, one study found widespread activations during NREM2, including in medial (BA 9) and lateral (BAs 10 and 46) prefrontal areas, anterior cingulate and insula, and a variety of subcortical and brainstem regions (Balkin et al., 2002). In stark contrast, another study found almost exclusively deactivations throughout the brain associated with NREM2, including in prefrontal cortex, inferior parietal lobule, superior temporal gyrus, insula, and various thalamic nuclei; indeed, only a single significant activation in the inferior frontal gyrus was observed (Kaufmann et al., 2006). Adding to this confusing picture, a study examining hippocampal functional connectivity throughout the brain found that there were no greater areas of connectivity for waking vs. NREM2, but, conversely, that a wide variety of regions showed increased coupling with hippocampus during NREM2 vs. waking, including many regions implicated in waking self-generated thought (Fox et al., 2015), such as the posterior cingulate cortex, lingual gyrus, inferior parietal lobule, temporopolar cortex, and insula (Andrade et al., 2011).

The neural correlates of NREM2 sleep therefore remain elusive and the limited data available difficult to synthesize. Part of the problem may be the intermediate (and by implication, highly variable) levels of cognitive activity and dreaming present in NREM2, which might be related to the similar variability of results observed in neuroimaging studies. That is, two given segments of sleep both scored as NREM2 based on relatively superficial similarities among EEG markers, and then pooled in neuroimaging analyses, might in fact be characterized by very different psychological content depending on the subjects recruited as well as the time of night (Cavallero, Cicogna, Natale, & Occhionero, 1992; Cicogna, Natale, Occhionero, & Bosinelli, 1998) and therefore might result in correspondingly distinctive patterns of brain recruitment (as observed in the studies to date). Further neuroimaging research into the neural basis of NREM2 accompanied by collection of mentation reports following laboratory awakenings would greatly help to clarify this situation.

*Neural correlates of NREM3/4 sleep (SWS)*

In contrast to NREM1 and NREM2, NREM3/4 (SWS) has been characterized mostly by widespread *deactivations* (relative to waking) throughout the brain (Andrade et al., 2011; Braun et al., 1997; Maquet et al., 1997; Maquet et al., 2005). For instance, Braun and colleagues (1997) found deactivations in SWS in prefrontal executive regions, such as the dorsolateral (BA 46) and ventrolateral (BA 11) prefrontal cortex, as well as in a variety of regions implicated in waking self-generated thought (Fox et al., 2015), including medial prefrontal cortex (BA 10), temporopolar cortex (BA 38), and anterior insula. Widespread deactivations were also observed in subcortical structures, including the basal ganglia, thalamus, and cerebellum (Braun et al., 1997). Similarly, Maquet and colleagues (1997) found SWS to be negatively correlated with regional cerebral blood flow in prefrontal areas such as orbitofrontal cortex (BA 11/25) and anterior cingulate cortex (BA 24), as well as some visual areas like precuneus (BA 19/7) and a variety of subcortical structures. Maquet and colleagues (2005) found further evidence for deactivations predominantly in executive, default, and visual areas. Finally, a connectivity study using hippocampus as the seed region found widespread reductions in functional connectivity during SWS as compared to waking, again primarily in prefrontal regions and default network areas such as the inferior parietal lobule, medial prefrontal cortex, and posterior cingulate cortex (Andrade et al., 2011).

The predominant pattern in SWS, based on the limited evidence to date, is deactivation (and/or disintegration of functional connectivity) throughout the brain – especially in prefrontal executive areas, default network regions, and subcortical structures.

*Neural correlates of REM sleep*

REM sleep has been associated with recruitment of widespread brain regions (Braun et al., 1997; Braun et al., 1998; Finelli et al., 2000; Maquet et al., 1996; Maquet et al., 2005; Peigneux et al., 2001). A sufficient number of studies have been conducted to allow for a preliminary meta-analysis of brain activations during REM sleep, which we recently executed in an effort to quantitatively assess the consistency of these activations (Fox et al., 2013). Combining our meta-analytic results with a qualitative assessment of individual studies, REM sleep appears to be associated with activation of the medial temporal lobe bilaterally; multiple regions within the default mode network, including clusters in medial prefrontal cortex (BA 24 and 9/32), dorsomedial prefrontal cortex (BA 9), and orbitofrontal cortex (BA 25); and numerous visual network areas, centered on the lingual gyrus (BA 18/19) (Domhoff & Fox, 2015; Fox et al., 2013). Deactivations are most salient in prefrontal executive regions, including dorsolateral (Braun et al., 1997; Braun et al., 1998; Maquet et al., 1996) and ventrolateral prefrontal cortex (Braun et al., 1997; Braun et al., 1998; Maquet et al., 2005).

*Neural correlates of lucid REM (LREM) sleep*

Only a single study to date has compared neural correlates of lucid REM sleep to regular REM sleep (Dresler et al., 2012), and this study relied on data from a single subject able to attain dream lucidity repeatedly in the scanner. Nonetheless, results of this pioneering (if tentative) study are intriguing. LREM, compared to standard, non-lucid REM sleep, was associated most notably with increased activation of default, visual, and frontoparietal control network regions (Dresler et al., 2012). These heightened activations are commensurate with the subjective qualities of LREM sleep, discussed further in the following sections.

## Discussion

*Default and visual network activation and their relationship to self-generated thought and imagery throughout the sleep cycle*

Dreaming can be thought of as an unconstrained, hyper-associative, and highly immersive form of self-generated thought that is largely detached from external sensory inputs (Christoff et al., 2016; Fox et al., 2013; Windt, 2010). We therefore might expect that areas involved in memory recall and recombination, self-referential thinking, and audiovisual imagery would show heightened recruitment compared to a restful waking baseline in all sleep stages that have some appreciable amount of thought or dream content (especially REM sleep, but also potentially stages NREM1 and NREM2).

Generally speaking, the NREM sleep stages are associated with considerably lower rates of dreaming than REM sleep (see Table 1) – and the mentation that does occur tends to be less visuospatial and immersive, and more of a conceptually-focused inner monologue (Hobson et al., 2000; Nielsen, 2000; Nir & Tononi, 2010). During NREM1, however, highly vivid visual imagery and occasionally full-blown (if short-lived) dreaming can occur (Hayashi et al., 1999; Hori et al., 1994; Nielsen, 1992; Stenstrom, Fox, Solomonova, & Nielsen, 2012). Consistent with these phenomenological reports, NREM1 is primarily associated with activation of secondary and tertiary visual regions, including the fusiform gyri bilaterally, the middle occipital gyrus (BA 19), and the cuneus (BA 18). Absent, however, are any notable activations in default network regions; this absence of activation, however, can be reconciled with the typical brevity and diminished sense of self (compared to other sleep stage mentation reports) that characterize sleep onset imagery (Cicogna et al., 1998).

Late in the night, when REM sleep predominates in the sleep cycle, NREM2 can give rise to a high frequency of immersive dream experiences indistinguishable from REM sleep reports (Cicogna et al., 1998). Cognitive activity of some kind is relatively frequent during early-night NREM2 (~50% of awakenings), but dreaming proper is more rare, reported from roughly 15-25% of NREM2 awakenings (Goodenough, Lewis, Shapiro, Jaret, & Sleser, 1965; Nielsen, 1999, 2000). As discussed above, however, the neuroimaging results concerning NREM2 are somewhat contradictory and do not lend themselves to any clear synthesis as of yet. Future work placing greater emphasis on first-person reports and sub-stage specificity may clarify this situation.

The frequency of dream experience is lowest by far in NREM3/4 (SWS) (Hobson et al., 2000; Nielsen, 2000). Indeed, significant numbers of participants can *never* recall cognitive activity or dreaming of any kind from SWS awakenings, despite multiple nights spent in the laboratory (Cavallero et al., 1992). Consistent with this very modest level of self-generated thought, SWS generally shows deactivation throughout major default network hubs, including medial prefrontal cortex and posterior cingulate cortex. Deactivations are also frequently reported in multiple subcortical brain areas, including the hypothalamus, thalamus, and pons. These subcortical deactivations are consistent with the overall decreased arousal and blockade of sensory inputs in SWS (Hobson et al., 2000).

REM, the sleep stage with by far the highest rates of dreaming (80-90% of the time (Hobson et al., 2000), shows heightened activation in numerous regions implicated in self-generated thought and imagery (Fox et al., 2015), especially widespread activation of the medial prefrontal cortex, the medial temporal lobe, and medial occipital areas (Domhoff & Fox, 2015; Fox et al., 2013). All of these activations are consistent with the endogenous generation of a self-referential narrative situated in a largely visual imaginal world.

These overall trends in activation patterns across the sleep stages are paralleled by changes in functional connectivity: connectivity among default mode network hubs, for instance, decreases monotonically throughout the NREM sleep stages (Sämann et al., 2011; Wilson et al., 2015). Consistent with these results, PET investigations have found a monotonic decrease in cerebral energy metabolism across NREM stages 1-3 (Maquet, 1995), whereas energy metabolism in REM sleep is equal to (Braun et al., 1997; Madsen et al., 1991; Maquet et al., 1990) or higher than (Buchsbaum et al., 1989; Heiss, Pawlik, Herholz, Wagner, & Wienhard, 1985) waking rest.

Finally, lucid REM sleep (LREM), in stark contrast to NREM sleep, shows activations *greater* even than non-lucid REM sleep in many regions. The most striking difference is the reappearance of activity in the frontoparietal control network, discussed in more detail in the following section. Also apparent is heightened activity in areas already hyperactive in REM sleep, including medial prefrontal cortex and a large swathe of medial occipitoparietal cortex – potentially explained by anecdotal reports that lucid REM sleep experiences are much more vivid and detailed than regular REM dreams (Dresler et al., 2012; Green, 1968; Sergio, 1988; Yuschak, 2006).

*Prefrontal executive deactivation, cognitive control, and meta-awareness throughout the sleep cycle*

As discussed elsewhere (Fox et al., 2015), waking self-generated thought involves a co-activation of default network areas alongside executive frontoparietal control network regions, most notably dorsal anterior cingulate cortex, rostrolateral prefrontal cortex, and anterior inferior parietal lobule. These results are not particularly difficult to rationalize when it is recalled that cognitive control and meta-awareness, the principal functions tied to the latter network, are in fact quite prevalent in waking self-generated thought (Christoff, Gordon, Smallwood, Smith, & Schooler, 2009; Christoff et al., 2016; Klinger, 1978, 2008; Klinger & Cox, 1987; Klinger & Kroll-Mensing, 1995; Kroll-Mensing, 1992; Seli, Risko, Smilek, & Schacter, 2016) – even if lower than in typical externally-directed cognition and tasks.

Conversely, executive and metacognitive functioning is largely absent or deficient in NREM and REM sleep cognition. Although dreams reports show strong thematic continuity with the emotional and personal concerns of waking life (Cartwright, Lloyd, Knight, & Trenholme, 1984; Fox et al., 2013; Kuiken, Dunn, & LoVerso, 2008), actual goal-related thinking or top-down control and planning are rare. Further, activities involving sustained top-down control of attention, such as reading, writing, or using a phone or computer, occur only very rarely in dreams (Hartmann, 1996; Schredl, 2000). Although logical and paralogical thinking indeed take place in sleep and dreaming (Hall & Van de Castle, 1966; Kahn & Hobson, 2005), overall such thinking is only peripherally goal-related at best and is deficient in many other respects (Kahn & Hobson, 2005).

Metacognitive functioning is similarly compromised. Natural rates of meta-awareness of one's true state during sleep (i.e., lucid dreaming or lucid sleep) are estimated to be only about 1%, even in experienced lucid dreamers (Schredl & Erlacher, 2004, 2011; Snyder & Gackenbach, 1988), and might occur only a handful of times throughout the entire lifespan in normal individuals (Barrett, 1991; Zadra, Donderi, & Pihl, 1992). Moreover, employing a variety of interventions in an effort to increase meta-awareness during sleep (including psychological training, pharmacological agents, and external electrical stimulation), even among highly-motivated participants, tends to result in only very modest and poorly-validated gains (Stumbrys, Erlacher, Schädlich, & Schredl, 2012). Meta-awareness of other features of experience, such as bizarre or impossible situations and discontinuities of time and place (Dorus, Dorus, & Rechtschaffen, 1971), is likewise heavily compromised (Kahn & Hobson, 2005).

Consistent with these many executive and metacognitive deficiencies throughout the sleep cycle, all sleep stages show at least some evidence for deactivation of prefrontal executive regions critical to cognitive control and meta-awareness. In stark contrast to these results throughout the rest of the sleep cycle, the unusual state of lucid REM (LREM) sleep instead shows *activation* of frontoparietal control network regions, including rostrolateral and dorsolateral prefrontal cortices, as well as in the anterior inferior parietal lobule bilaterally – consistent with restored cognitive control and meta-awareness (Dresler et al., 2012; Fox & Christoff, 2014).

*Limitations*

Four major limitations of our discussion should be emphasized. The first is that the subjective reports of nighttime dreaming and cognition have been collected largely independently of the functional neuroimaging data that speaks to brain recruitment during the various sleep stages. That is, although differences in subjective experience across sleep stages are in general reliable and well-replicated (Hobson et al., 2000; Nielsen, 1999, 2000), the studies that have used functional neuroimaging to examine these same sleep stages have rarely actually collected dream or mentation reports from their participants (for instance, of the eight studies we reviewed of REM sleep, only one confirmed dreaming had indeed been taking place in the REM sleep periods examined in the PET scanner: (Maquet et al., 1996)). In the absence of such reports collected directly following functional neuroimaging of sleep and dreaming, any putative relationship between sleep stage neurophysiology and subjective content remains, at best, inferential and probabilistic. The obvious solution to this problem is for future functional neuroimaging studies of sleep to actively awaken and interrogate participants as to their subjective experiences across various sleep stages. The putative links between subjective experiences and brain activation discussed here are therefore almost entirely inferential and based on statistical averages of content reports generated in independent studies that did not use PET or MRI scans to assess brain activation. The apparent linkages we speculate on here are therefore, at best, crude approximations of average experiential intensity and average brain activation across many different subjects; although intriguing, these correspondences need to be further explored and corroborated with more detailed and targeted research before they can be considered reliable, much less definitive.

A second major concern is the small number of studies that have so far investigated any given sleep stage. Although we took pains to search the literature thoroughly and review every well-controlled and rigorously executed study, nonetheless the field of neuroimaging of sleep remains small. Accumulating research in this domain, however, will gradually mitigate this concern, as more powerful and representative syntheses become possible.

Third is the fact that a full half of all studies we consulted sleep-deprived their participants the night before brain scanning in order to facilitate the maximum amount of sleep in the scanner. Aside from nonspecific effects, such as stress, sleep deprivation is known to affect the architecture and EEG correlates of the sleep cycle (Borbély, Baumann, Brandeis, Strauch, & Lehmann, 1981), and might therefore influence neuroimaging measures of brain recruitment as well. In principle this concern could be addressed by comparing brain activation and deactivation for given sleep stages across studies that did and did not employ sleep deprivation, but a much larger body of research is required before any such comparison is possible.

A fourth and final concern that should be reiterated is that sleep stages are probably as much a convenient abstraction as they are a concrete neurophysiological fact. That is, while not strictly categorical, general differentiation between distinct sleep stages is justifiable and represents a valid and useful explanatory tool. We refer the reader back to the introductory material for a more detailed consideration of this important issue.

*Conclusions*

In a major review more than 15 years ago, Hobson et al. (2000) defended the position that REM and NREM sleep stages *"can* be defined, that their components can be analyzed and measured, and that they *are* significantly different from one another" (original emphasis). We have here provided some new evidence in favor of this view from an overview of functional neuroimaging studies across every stage of the sleep cycle. We have presented preliminary evidence that subjective experiences and neurophysiological markers co-vary across the sleep cycle, with the most intriguing finding being that DMN and visual network activation might track the occurrence and immersiveness of self-generated thought, whereas frontoparietal control network deactivation might track the general loss of cognitive control over, and meta-awareness of, one's psychological state. Overall, these results provide an intriguing example of complex, but nonetheless coherent, patterns of brain-mind isomorphism (Cacioppo & Tassinary, 1990). We hope that this tentative synthesis will serve as a useful stepping-stone on the path to a much deeper understanding of sleep and dream neurophysiology that surmounts the major limitations inherent in present research.